\documentclass[a4paper]{article}

\usepackage{LaTeX/INTERSPEECH2021}
\usepackage{caption}
\usepackage{subcaption}
\usepackage{multicol}
\usepackage[symbol]{footmisc} 
\usepackage{multirow} 
\usepackage{xspace}

\title{ADEPT: A Dataset for Evaluating Prosody Transfer}
\name{
    Alexandra Torresquintero$^1$,
    Tian Huey Teh$^1$,
    Christopher G. R. Wallis$^1$,
    Marlene Staib$^1$, \\
    Devang S Ram Mohan$^1$,
    Vivian Hu$^1$,
    Lorenzo Foglianti$^1$,
    Jiameng Gao$^1$,
    Simon King$^{1,2}$}
\address{
  $^1$Papercup Technologies Ltd., United Kingdom \\
  $^2$University of Edinburgh, United Kingdom}
\email{\{alexandra,tian,chris\}@papercup.com}

\def\F0{$F_{0}$\xspace}

\begin{document}

\maketitle


\vspace{-1mm}
\begin{abstract}
\vspace{-1mm}

Text-to-speech is now able to achieve near-human naturalness and research focus has shifted to increasing expressivity. One popular method is to transfer the prosody from a reference speech sample. There have been considerable advances in using prosody transfer to generate more expressive speech, but the field lacks a clear definition of what successful prosody transfer means and a method for measuring it.
We introduce a dataset of prosodically-varied reference natural speech samples for evaluating prosody transfer. The samples include global variations reflecting emotion and interpersonal attitude, and local variations reflecting topical emphasis, propositional attitude, syntactic phrasing and marked tonicity. The corpus only includes prosodic variations that listeners are able to distinguish with reasonable accuracy, and we report these figures as a benchmark against which text-to-speech prosody transfer can be compared.
We conclude the paper with a demonstration of our proposed evaluation methodology, using the corpus to evaluate two text-to-speech models that perform prosody transfer.
\end{abstract}

\vspace{-1.5mm}
\section{Introduction}
\vspace{-0.5mm}
Text-to-speech (TTS) research relies on subjective human evaluations. There are well-established methods to assess naturalness -- A/B comparisons, Mean Opinion Score (MOS), or MUSHRA \cite{MUSHRA} -- and intelligibility using a transcription task.
But, now that TTS routinely matches human speech intelligibility and approaches human naturalness, focus has shifted to expressivity. Perhaps the most popular current method is to transfer prosody from an expressive reference sample. 

Subjective methods for evaluating prosody transfer are not well developed. Some provide listeners with a reference and measure how well its prosody was transferred. \cite{pmlr-v80-skerry-ryan18a, 1906.03402, 1911.09645} employed an AXY discrimination task in which listeners judge how similar generated samples X and Y are to reference A; \cite{Karlapati2020} used this task with trained linguists. \cite{lee_2019} claimed that successful transfer of a song melody to speech indicates successful prosody transfer.
Other methods provide no reference. \cite{Kenter2020} use a preference test.
\cite{1906.03402, 1911.09645, sun} assume that measuring naturalness with MOS is sufficient.
%
There is no common method, which hinders the comparison of prosody transfer approaches. Therefore, we release a corpus (DOI: 10.5281/zenodo.5117102) of expressive natural speech reference samples that can be used within our proposed evaluation methodology.

The goal of prosody transfer is to synthesise a given text with the prosody of a reference utterance. By varying the reference, the system generates prosodically-distinct renditions. Our proposed evaluation method therefore starts from a corpus of English sentences, each with multiple prosodically-distinct natural renditions. The TTS system under evaluation is required to transfer the prosody from a reference utterance taken from this corpus, and the evaluation metric is the accuracy with which listeners perceive the correct prosody. 

To identify the effectiveness of the transfer across different aspects of prosody, the natural utterances fall into prosodic classes within which listeners are able to perform a categorisation task. To find suitable classes, we examine several spoken phenomena known to have prosodic consequences (§\ref{sec-categories}). Within each of these, we identify subcategories reported in the literature to have perceptually distinct prosody (§\ref{sec-design-subcategories}). The listeners' task will then be to categorise a synthetic rendition as the correct subcategory.
After recording the natural utterances (§\ref{sec-design-recordings}), we confirm that listeners can perform the  categorisation task on natural speech (§\ref{sec-pretests}), and discard subcategories and utterances that listeners cannot reliably categorise.
We finish with a demonstration of our proposed method using synthetic speech (§\ref{sec-results}), and a discussion and suggestions for further work (§\ref{discussion}).

\vspace{-1.5mm}
\section{Speech classes with prosodic effect} \vspace{-0.5mm}
\label{sec-categories}
Prosody has many definitions, but we adopt \cite[p. 196]{shattuck-hufnagel}: high-level structures that account for \F0, duration, amplitude, spectral tilt, and segmental reduction patterns in speech. These structures have local and global effects \cite{Moraes-2011}. Many aspects of speech are part of prosody by this definition. 



\cite{Moraes-2011} describes two phenomena, emotion and attitude, that speakers express through $F_{0}$, amplitude, duration, and spectral tilt prosodic cues \cite{Moraes-2011, Rillard}. \textbf{Emotion} is an inner state of the speaker (e.g., joy), whilst attitude is towards something external. \textbf{Interpersonal attitude} is toward the listener, e.g., friendliness. \textbf{Propositional attitude} is toward what is being said, e.g., incredulity. Emotion and interpersonal attitude have a global prosodic effect, and propositional attitude has a local effect \cite{Moraes-2011}.

\textbf{Topical emphasis} occurs when the topic is prosodically highlighted because of its relative importance to other words in the sentence, such as \emph{not} in \emph{I will NOT go}. It has local effects on \F0, amplitude, and duration \cite[p. 15]{BOLINGER}.

\textbf{Syntactic phrasing} affects prosody through perceivable intonation groups: the end of a phrase exhibits lengthening of the phrase-final word and following pause within a sentence \cite{Allbritton}.

In English speech there will always be a syllable that carries the greatest lexical stress across the sentence \cite{mateo}\cite{Wells}; we call this phenomenon \textbf{marked tonicity}. It has similar, though subtler, prosodic effect on prominent words as topical emphasis, but can also cause segmental reduction on non-stressed words.

We have introduced 2 classes that have a global prosodic effect (emotion and interpersonal attitude), and 4 classes that have local effect (propositional attitude, topical emphasis, syntactic phrasing, and marked tonicity). Other structures have similar prosodic effects, such as style (whispered, instructional, broadcasting, etc.) and speaker identity (age, gender, accent, etc.). However, these are less likely to change per-utterance, so are less relevant to most TTS prosody transfer use cases.

\section{Dataset and evaluation design}
\label{sec-design}
These 6 classes are not mutually exclusive; in one sentence a speaker can sound sad (emotion) and polite (interpersonal attitude) and emphasise a word. But to use these classes to evaluate TTS prosody transfer, we require that their prosodic effects are perceivable in isolation. We propose a disambiguation task in which listeners are asked to categorise speech samples based only on their prosody. For example, the sentence \emph{It's snowing} can be said both happily or sadly. If listeners who are played both recordings can correctly identify which is sad and which is happy, we can conclude listeners can perceive emotion based on prosody alone. In the following sections, we describe how we used prior research to determine suitable ambiguity for such a disambiguation task (§\ref{sec-design-subcategories}), the design and recording of the natural speech from which prosody will be transferred (§\ref{sec-design-recordings}), pretests to find the most reliable task design and data for evaluating prosody transfer (§\ref{sec-pretests}), and our final proposed evaluation methodology (§\ref{section:final_methodology}).

\subsection{Perceivable subcategories or interpretations}
\label{sec-design-subcategories}
For each of the 6 classes, listeners will perform a disambiguation task using prosody: therefore, we
needed to identify prosodic ambiguity within each class. 
For emotion, interpersonal attitude, propositional attitude, and topical emphasis classes, we found suitable ambiguity in \emph{subcategories} of the class.
For syntactic phrasing and marked tonicity, 
sentences with two \emph{interpretations} had suitable ambiguity.

For \textbf{emotion},  \cite{Pell2009FactorsIT} report the perceivability of anger, disgust, fear, sadness, happiness, pleasant surprise, and neutral, in 4 languages. We discarded pleasant surprise as it was 1 in 3 of their perceptually-invalid items. We renamed happiness to joy which is less likely to be confused for something more complex like nostalgia. This left 5 perceptually distinct subcategories of the emotion class: anger, disgust, fear, sadness, and joy.

\cite{Moraes-declarative, Moraes-interrogative} measured the perceivability (in Brazilian Portuguese) of 12 \textbf{interpersonal attitudes}. Listeners had to disambiguate arrogance, authority, contempt, irritation, politeness, seduction, and neutral in question and statement utterances. We eliminated subcategories whose perceivability interacted with speaker gender (arrogance, irritation, seduction), and subcategories confused with neutral (contempt statements, polite questions). We eliminated authoritative questions because authoritative statements were perceived more strongly. This left three subcategories of the interpersonal attitude class: contemptuous questions, authoritative statements, and polite statements.

\cite{Moraes-declarative, Moraes-interrogative} also measured \textbf{propositional attitude} by asking listeners to disambiguate between four question attitudes (rhetoricity, confirmation, incredulity, surprise) plus neutral, and five statement attitudes (irony, incredulity, surprise, doubt, obviousness) plus neutral. We eliminated rhetoricity questions for being confused with neutral, surprise questions for being confused with incredulity, and incredulity statements for being confused with irony. We renamed irony to sarcasm for clarity. This left 6 subcategories of the propositional attitude class: obviousness, surprise, sarcasm, and doubt statements, and incredulity and confirmation questions. 

\cite{eady} show that acoustic differences can arise depending on the locus of \textbf{topical emphasis} in the sentence, yielding three subcategories: beginning, middle, and end.

\cite{Kraljic} show that listeners can use prosodic cues to disambiguate meanings of a sentence with phrasing ambiguity. For example, \emph{Put the dog food in the bowl on the floor} has two interpretations: put the dog food into the bowl that is on the floor, or put the dog food that is in the bowl onto the floor. The former meaning can be conveyed with a pause after \emph{food}, and the latter with a pause after \emph{bowl}. We used sentences with this \textbf{syntactic phrasing} ambiguity for our disambiguation task. These are not subcategories per se, so we refer to them as interpretations.

Sentences with part of speech ambiguity can be disambiguated by \textbf{marked tonicity} prosodic cues. \cite[p. 55]{Hirst} gives an example sentence \emph{He ate a little pudding} which has two interpretations:
1) he didn't eat very much pudding, in which `a little' is a determiner to `pudding', and the strongest lexical stress falls on `pudding'; 2) he ate a small pudding, where it falls on `little', which is an adjective for `pudding'.

For each of the 6 classes, the next step was to design sentences that can be read in prosodically distinct ways by subcategory (of emotion, interpersonal attitude, propositional attitude, topic emphasis) or interpretation (for syntactic phrasing, marked tonicity), and record them with voice actors.



\subsection{Sentence design and recording}
\label{sec-design-recordings}
For each class, we devised at least 20 sentences that could be spoken to express all the subcategories of, or two interpretations per sentence within, that class. For example, \emph{``Look at that puppy.''}, can be said in all five subcategories of the emotion class. Many of the sentences for syntactic phrasing and marked tonicity came from \cite{price}.

In addition to the sentence to be spoken, we devised a contextual cue to elicit the prosodically-distinct rendition portraying each subcategory/interpretation.
For \textbf{emotion}, \textbf{interpersonal attitude}, and \textbf{propositional attitude}, the contextual cues were situations that evoke the subcategory, such as a teacher-student relationship implying an authoritative interpersonal attitude.
For \textbf{topical emphasis}, the emphasised word was capitalised in the script, and contextual cues were wh-questions that implied the emphasis. For example, for the sentence \emph{``Dogs play FETCH in parks''} with target emphasis on FETCH, the contextual wh-question was ``Dogs play WHAT in parks?''
For \textbf{syntactic phrasing} and \textbf{marked tonicity}, the contextual cues were paraphrases that made the intended meaning clear, such as the two explicit interpretations of the sentence \emph{Put the dog food in the bowl on the floor} in §\ref{sec-design-subcategories}.
When possible, each sentence was also recorded in a `neutral' style: no subcategory was expressed. As one of the two interpretations is required for each syntactic phrasing and marked tonicity sentence, a neutral style for these classes does not exist. Two voice actors, male and female, read the 552 sentences in Standard Southern British English.

\subsection{Pretests}
\label{sec-pretests}
Before using these sentences and subcategories/interpretations to evaluate prosody transfer, we tested that listeners consistently found them prosodically distinguishable in natural speech, and then selected the most appropriate task for them to perform.

We trialled various disambiguation task designs for each class (§\ref{sec-pretest-task-design}), then selected the most reliable design, and finally used that task to select the most consistently disambiguated sentences and subcategories for each speaker/class pair (§\ref{sec-pretest-elimination}). The resulting task design and selected sentences will be used in the final evaluation methodology (§\ref{section:final_methodology}).



\subsubsection{Listeners}
\label{section:listeners}
Amazon Mechanical Turk was used to recruit 10 self-reported native English speakers who passed a short transcription test, per pretest.
To filter listeners who could hear prosodic differences, we disqualified  participants who:
1) across all questions, selected an option significantly above chance (e.g., always selecting A);
or 2) for a given correct subcategory, did not select any subcategory significantly above chance.
E.g., for all questions whose correct answer is disgust, people who statistically significantly
selected angry (or any other subcategory) were not disqualified.
As 10 is a small sample size, we treated answers as binomial distributions with statistical significance at the 99\% confidence interval.

\begin{figure}
    \centering
    \begin{subfigure}[b]{0.5\columnwidth}
        \centering
        \caption{Emotion (Female)}
        \vspace{-0.6\baselineskip}
        \includegraphics[width=\columnwidth]{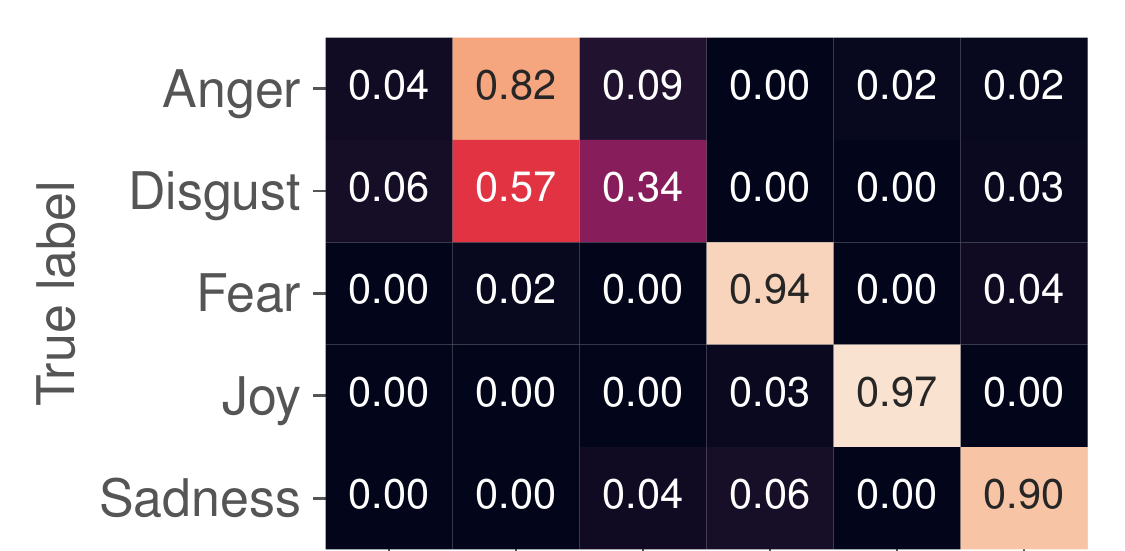}
        \vspace{-1.25\baselineskip}
    \end{subfigure}%
    \hfill
    \begin{subfigure}[b]{0.5\columnwidth}
        \centering
        \caption{Propositional Attitude (Female)}
        \vspace{-0.4\baselineskip}
        \includegraphics[width=\columnwidth]{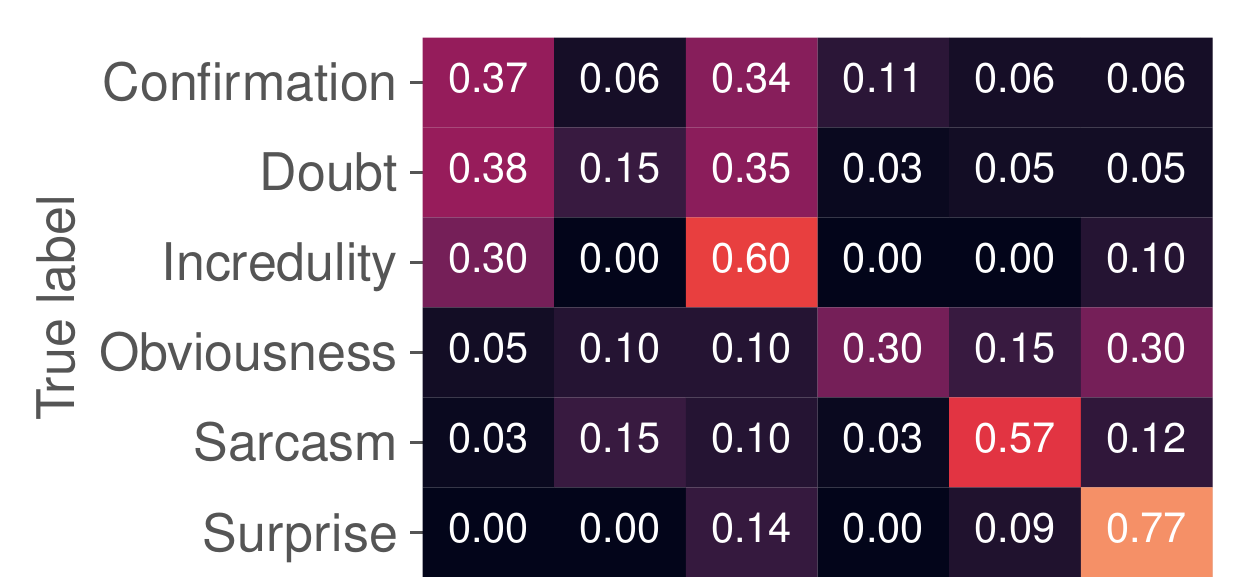}
        \vspace{-1.1\baselineskip}
    \end{subfigure}
    \begin{subfigure}[b]{0.5\columnwidth}    
        \centering
        \includegraphics[width=\columnwidth]{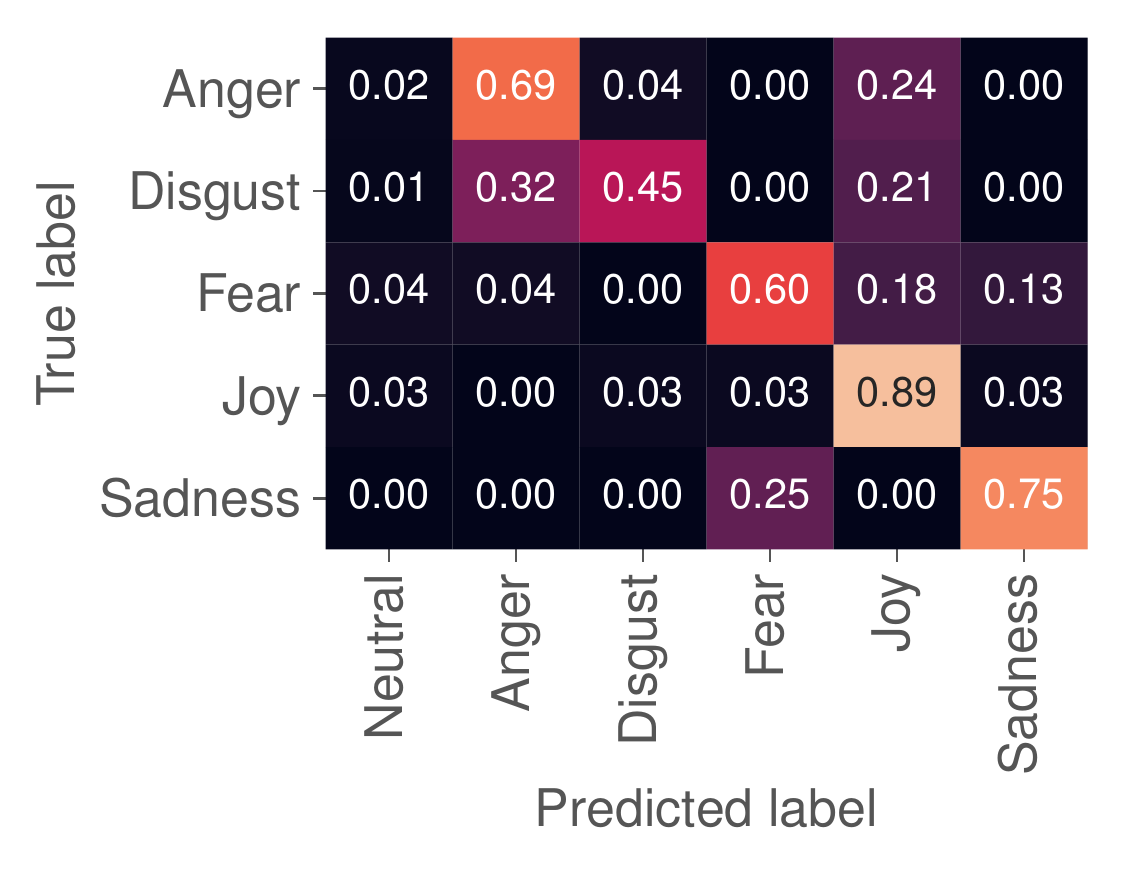}
        \vspace{-1.5\baselineskip}
        \caption{Emotion (Male)}
    \end{subfigure}%
    \hfill
    \begin{subfigure}[b]{0.5\columnwidth}    
        \centering
        \includegraphics[width=\columnwidth]{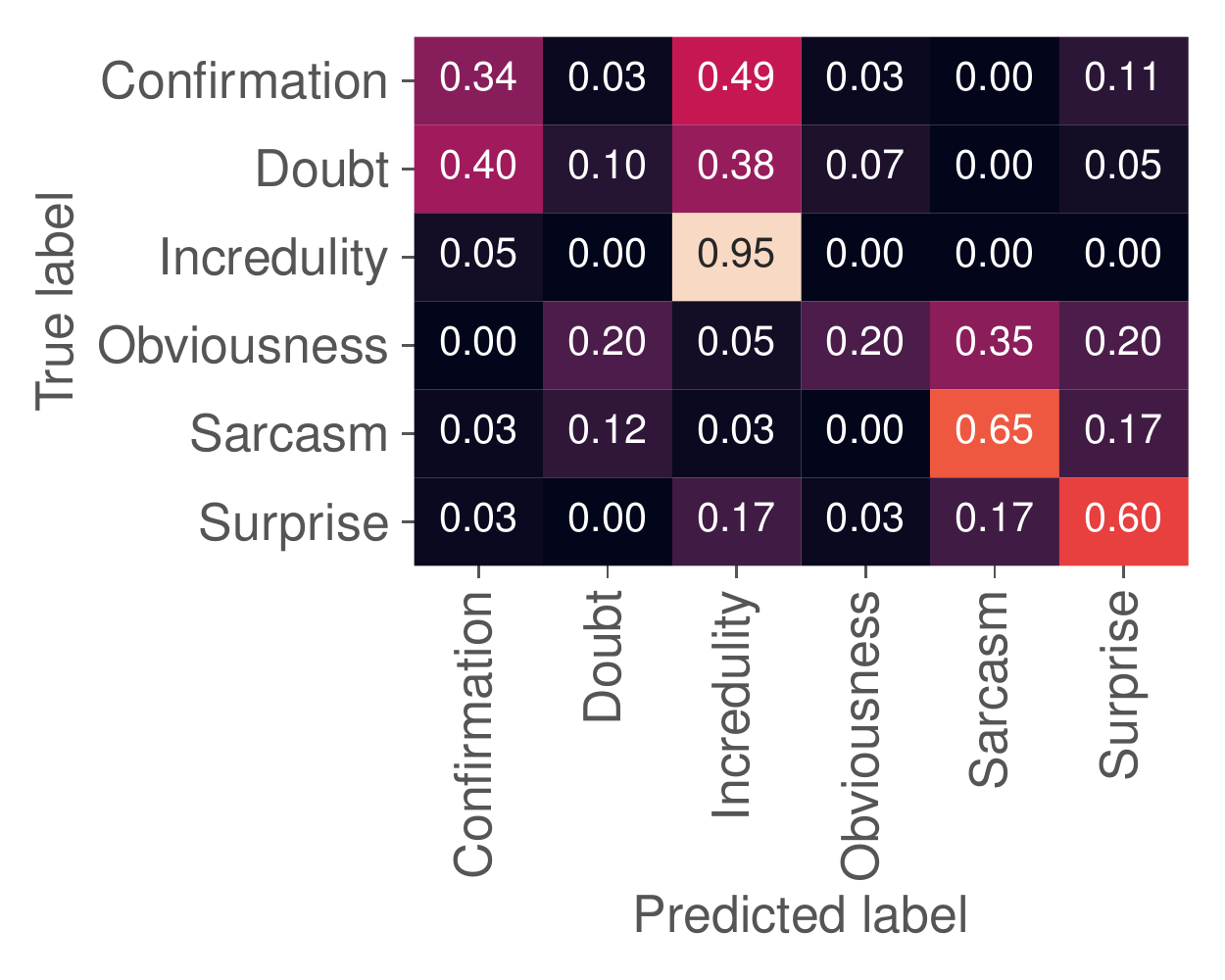}
        \vspace{-1.5\baselineskip}
        \caption{Propositional Attitude (Male)}
    \end{subfigure}
    \vspace{-5mm}
    \caption{Confusion matrices for the top 5 sentences from classes with subcategories eliminated in pretests.}
    \label{fig:pretest_confusion_matrices}
    \vspace{-1\baselineskip}
    \vspace{-2mm}
\end{figure}

\vspace{-0.5mm}
\subsubsection{Task design}
\label{sec-pretest-task-design}
As with the design of any listening evaluation, there were more potential design options than could be fully explored. Because we would subsequently be eliminating weak stimuli (§\ref{sec-pretest-elimination}) which would further improve recognition accuracy, in this task design phase we used our best judgement to make design decisions, but acknowledge that this will not always be optimal.

We assume task design is speaker-independent, so trialled multiple task designs only on the female speaker. The neutral stimulus, when it existed, was included as a sample unless otherwise specified. We considered two design choices: 1) Should we ask \textit{directly} or \textit{indirectly} about the subcategory/interpretation? E.g., do we directly ask which word carries most lexical stress, or present the sample in a context where that stress pattern would be preferred?
2) Should listeners categorise a \textit{single stimulus} or choose which one of \textit{multiple stimuli} matches a label?




For \textbf{emotion}, we tried a direct question. 
The multiple stimulus design asked listeners `In which of the following samples does the speaker sound most \emph{x}?' where \emph{x} was one of disgusted, angry, joyful, fearful, or sad. The stimuli were samples of the same sentence in each of the subcategories plus neutral.
The single stimulus design asked listeners `Which emotion is most reflected in the speakers voice?', with six choices: disgust, anger, joy, fear, sadness, or none of these.
Across all 20 sentences, we found higher recognition accuracy for each subcategory in the multiple stimulus task.


For \textbf{interpersonal attitude}, we tried a direct question multiple stimulus design. We asked `Which of the following samples sounds most like \emph{x}?' where \emph{x} was an authoritative statement, a contemptuous question, or a polite statement.

For \textbf{topical emphasis}, we tried a single stimulus design and compared direct and indirect questions. The direct question asked listeners `Which word is most strongly emphasised in the sample?', with three choices of the content words in the beginning, middle, and end of the sentence. The indirect question asked `Which question is best answered by the sample?' The three choices were the context cues described in §\ref{sec-design-recordings}. Neutral was not included because this is not a correct response to any of the context cues. After disqualifying participants, recognition accuracies across all sentences for each subcategory were higher for the indirect design.

We tried an indirect question multiple stimulus design for \textbf{propositional attitude}, but excluded neutral  because we believed 7 samples (6 subcategories $+$ neutral) was too many to compare at once. Participants were asked `Which audio fits best into the context: \emph{x}?' where \emph{x} was one of:

\indent\indent\indent``Obviously \underline{\hspace{1cm}}.''

\indent\indent\indent``(Surprised) Wow! \underline{\hspace{1cm}}!''

\indent\indent\indent``(Sarcastically) Well \underline{\hspace{1cm}}.''

\indent\indent\indent``(Unsure) Perhaps \underline{\hspace{1cm}}."

\indent\indent\indent``Really? \underline{\hspace{1cm}}?''

\indent\indent\indent``I just want to confirm that \underline{\hspace{1cm}}?''

For \textbf{syntactic phrasing}, we tried single stimulus and multiple stimulus designs with an indirect question. Listeners were asked `Which is a better paraphrase of the sample?' in the single stimulus design, and `Which sample fits the paraphrase best? \emph{x}' in the multiple stimulus design, where \emph{x} was a paraphrase from §\ref{sec-design-recordings}. The single stimulus design provided better results, potentially because it is easier to see the two alternative interpretations by reading two paraphrases. We also used this paraphrase single stimulus design for \textbf{marked tonicity}.

\vspace{-0.5mm}
\subsubsection{Elimination of weak stimuli}
\label{sec-pretest-elimination}
The best designs above for each class were re-run using male speaker stimuli. We then used the qualified participants to identify the five sentences for each speaker and class that had the highest recognition accuracy across all subcategories/interpretations.
We eliminated any subcategories with less than 60\% recognition accuracy in these sentences, assuming anything below this threshold was not perceivable enough to measure TTS against in the final evaluation method (§\ref{section:final_methodology}).

Disgust was discarded as a subcategory of \textbf{emotion} because its recognition accuracy was less than 60\% for both speakers (Figures \ref{fig:pretest_confusion_matrices}a and c).
All \textbf{interpersonal attitude} (Table \ref{tab:00-int-att-pre}) and \textbf{topical emphasis} subcategories met the 60\% threshold.
\begin{table}[h]
\vspace{-2mm}
  \caption{Interpersonal attitude pretest results}
\vspace{-2mm}
  \label{tab:00-int-att-pre}
  \centering
  \begin{tabular}{ c c c c }
    & authority & contempt & politeness \\
    \hline
    female & 60\% & 83\% & 85\% \\
    male & 93\% & 60\% & 71\% \\
  \end{tabular}
  \vspace{-4mm}
\end{table}
For topical emphasis, recognition accuracy was 100\% for all subcategories and speakers.
Figures \ref{fig:pretest_confusion_matrices}b and d show that confirmation, doubt, and obviousness \textbf{propositional attitudes} did not meet the 60\% threshold for either speaker, nor did sarcasm for the female speaker.
For the 2 classes with sentence-dependent interpretations, we report accuracy per speaker for all top 5 sentence stimuli together, because subcategories of these sentences do not exist. For \textbf{syntactic phrasing} this accuracy was 90\% for both speakers. For \textbf{marked tonicity}, this was 79\% and 83\% for the female and male stimuli respectively.

\subsection{The proposed evaluation methodology}
\label{section:final_methodology}
The final ADEPT evaluation methodology consists of 12 disambiguation tasks: 6 classes $\times$ 2 speakers. Each task uses 5 sentences with multiple prosodic renditions. Each task has one question per sentence and distinguishable subcategory or interpretation. As shown in Table \ref{tab:setup}, for each question there is one choice per subcategory or interpretation, plus neutral if it is present.
For example, the final female propositional attitude test is a 10 question multiple stimulus task, and each question's three choices are the incredulity, surprise, and neutral samples.

For the final setup for propositional attitude, we include neutral because some subcategories were eliminated, and the incredulity context is updated to ``(Incredulous) Really? \underline{\hspace{0.94cm}}?"


\begin{table}[t]
\centering
\caption{Number of choices per question for each disambiguation task, whether questions are single or multiple stimulus, and whether neutral samples are included as stimuli.}
\label{tab:setup}
\vspace{-2mm}
\begin{tabular}{ c | c c | c | c}
 & \multicolumn{2}{c|}{choices} & audio & neutral \\
 class & F & M & stimuli & included \\
 \hline
 emotion & 5 & 5 & multiple & yes \\

 interpersonal attitude & 4 & 4 & multiple & yes \\

 topical emphasis & 3 & 3 & single & no \\

 propositional attitude & 3 & 4 & multiple & yes \\

 synctactic phrasing & 2 & 2 & single & - \\

 marked tonicity & 2 & 2 & single & - \\
\end{tabular}
\vspace{-6mm}
\end{table}

\section{Evaluating TTS prosody transfer models} \label{sec-results}
We demonstrate the use of the ADEPT evaluation 
methodology to compare synthetic speech generated by two recently-proposed TTS models that perform prosody transfer. At the same time, we establish a benchmark based on the recognition accuracy of
natural speech.

Our two models are both based on a multi-speaker variant of Tacotron 2 \cite{tacotron2, multispeaker}. Following \cite{pmlr-v80-skerry-ryan18a} we extend the Tacotron 2 architecture by adding a reference encoder that learns a fixed-length prosody embedding from the reference in unsupervised fashion (henceforth \textbf{Tacotron-Ref}). We compare this to \textbf{Ctrl-P} \cite{Ctrl-P}, which explicitly models three acoustic correlates of prosody (\F0, energy, and duration) per-phone. For supervised training, ground-truth values are extracted from the force-aligned training data. Each feature is normalised to zero mean and unit standard deviation, per speaker. During inference, values are extracted from the reference speech and normalised using the target speaker's train set statistics. This variable-length prosody representation is concatenated with the Tacotron 2 encoder output and attended over by the decoder.

Our training data comprised 24~h of non-fiction audiobook readings by the female speaker from the LJSpeech corpus \cite{lj}, 20~h of fiction audiobook readings by the female speaker from the 2013 Blizzard Challenge \cite{blizzard}, and 3~h of proprietary data (not from ADEPT) in order to include a male speaker.

We trained one \textbf{Ctrl-P} model and one \textbf{Tacotron-Ref} model on this corpus. Female samples were generated with LJ as the target speaker with prosody transferred from samples of the female ADEPT speaker. Male samples were generated with our proprietary speaker as the target and prosody transferred from samples of the male ADEPT speaker. The Griffin-Lim \cite{griffin} algorithm was assumed to be sufficient for the current demonstration of the ADEPT evaluation methodology, but neural-vocoded samples could also have been used.

The evaluations on the natural speech and generated samples were performed on Amazon Mechanical Turk by self-reported native English speakers who passed a short English transcription test. In the pretests (§\ref{section:listeners}) using only natural speech, we disqualified participants who couldn't hear any prosody. Since TTS cannot guarantee successful prosody transfer, this is no longer appropriate. Instead, we employed 5 `trap' questions. For multiple stimulus designs, trap questions required the listener to identify the sample that sounded most like `English speech', where all audio files but one were time-reversed. For the single stimulus designs, trap questions replaced all options except the correct one with a description obviously unrelated to the sample. Participants who got any trap question wrong were disqualified and excluded from results. Each test had exactly 30 qualifying participants, as recommended by \cite{Wester}. Results are shown in Table \ref{tab:results},
with the 2 global classes above and the 4 local classes below.

\begin{table}[t]
\centering
\caption{Recognition accuracy (\%) for each of a class' subcategories, or entire class that doesn't have subcategories, for female (F) and male (M) natural speech (N), and LJSpeech (LJ) and proprietary (Prop.) synthetic speakers from Ctrl-P (C) and Tacotron-Ref (T). Accuracies statistically significantly above chance are in bold (one-tailed binomial test; $p\leq0.05$).}
\label{tab:results}
\vspace{-2mm}
\begin{tabular}{c c|c|c|c|c|c|c }
{} & {} & F & \multicolumn{2}{c|}{LJ} & M & \multicolumn{2}{c}{Prop.} \\
\hline
class & subcategory & N & C & T & N & C & T\\
\hline\hline
\multirow{4}{*}{emotion} & anger & \textbf{95} & \textbf{31} & 17 & \textbf{83} & 6 & 6 \\
 & fear & \textbf{80} & 16 & \textbf{40} & \textbf{52} & 20 & 9 \\
 & joy & \textbf{90} & \textbf{33} & 18 & \textbf{88} & \textbf{75} & \textbf{54} \\
 & sadness & \textbf{88} & \textbf{49} & 21 & \textbf{62} & \textbf{53} & 13 \\
\hline
inter- & authority & \textbf{47} & 14 & 26 & \textbf{60} & \textbf{35} & 29 \\
personal & contempt & \textbf{49} & \textbf{50} & 29 & \textbf{52} & \textbf{35} & 30 \\
attitude & politeness & \textbf{37} & 23 & 26 & \textbf{63} & \textbf{37} & 29 \\
\hline
\multirow{3}{*}{\shortstack{topical\\emphasis}} & beginning & \textbf{87} & \textbf{68} & \textbf{52} & \textbf{82} & \textbf{87} & \textbf{66} \\
 & middle & \textbf{79} & \textbf{67} & \textbf{43} & \textbf{69} & \textbf{71} & 33 \\
 & end & \textbf{70} & \textbf{67} & \textbf{45} & \textbf{62} & \textbf{63} & 32 \\
\hline
propo- & incredulity & \textbf{63} & 40 & 40 & \textbf{71} & \textbf{75} & \textbf{33} \\
sitional & sarcasm & - & - & - & \textbf{62} & \textbf{48} & \textbf{49} \\
attitude & surprise & \textbf{73} & 32 & 33 & \textbf{66} & \textbf{72} & \textbf{47} \\
\hline
\multirow{2}{*}{\shortstack{syntactic\\phrasing}} & & \multirow{2}{*}{\textbf{84}} & \multirow{2}{*}{\textbf{77}} & \multirow{2}{*}{\textbf{66}} & \multirow{2}{*}{\textbf{80}} & \multirow{2}{*}{\textbf{84}} & \multirow{2}{*}{\textbf{81}} \\
 & & & & & & & \\
\hline
\multirow{2}{*}{\shortstack{marked\\tonicity}} & & \multirow{2}{*}{\textbf{74}} & \multirow{2}{*}{\textbf{69}} & \multirow{2}{*}{\textbf{58}} & \multirow{2}{*}{\textbf{62}} & \multirow{2}{*}{\textbf{58}} & \multirow{2}{*}{48} \\
 & & & & & & & \\
\end{tabular}
\vspace{-6.5mm}
\end{table}

The ADEPT evaluation measures the success of prosody transfer for each model-voice combination on each subcategory, or class for classes without subcategories. Both Ctrl-P and Tacotron-Ref perform better than chance in some cases, with Ctrl-P doing so more and matching the natural speech benchmark for several classes. 

\vspace{-2mm}
\section{Discussion and Conclusion} \label{discussion}
As expected, accuracies for natural speech are all significantly above chance, albeit lower than in pretests, probably as a result of different qualifying criteria.
Beyond model comparison, the ADEPT evaluation methodology also enables a host of other analyses. For instance, one could investigate differences in model performance between local and global prosodic classes, examine within-class confusion matrices, or compare performance transferring prosody from different source voices.


In this work, we introduced six high level local and global prosodic classes of speech that can be used in disambiguation tasks to evaluate TTS prosody transfer. This evaluation methodology allows researchers to both compare performance of their models against each other, and against a natural benchmark of target performance. Further work might consider if these classes and their subcategories/interpretations are viable for a cross-lingual prosody transfer application, or if they can be used to evaluate prosody in TTS in general.

\section{Acknowledgements}
We thank our adviser Mark Gales for guidance and our voice actors Nishad and Laura who read all sentences even when they didn't make sense.

\bibliographystyle{IEEEtran}

\bibliography{main}
\end{document}